\documentclass[reprint,aps,pre,twocolumn]{revtex4-1}

\usepackage{ulem}
\usepackage{amsmath}
\usepackage{amsthm}
\usepackage{amssymb}
\usepackage{appendix}
\usepackage{blkarray}
\usepackage{bm}
\usepackage{booktabs}
\usepackage{calc}
\usepackage{dcolumn}
\usepackage{float}
\usepackage{graphicx}
\usepackage{listings}
\usepackage{mathtools}
\usepackage{setspace}
\usepackage[caption=false]{subfig}
\usepackage{verbatim}

\begin{document}

\preprint{APS/123-QED}

\title{Truncation selection and diffusion on lattices}

\author{Bryce Morsky}
\affiliation{Department of Mathematics and Statistics, 50 Stone Road East, Room 437 MacNaughton Building, University of Guelph, Guelph, Ontario, Canada N1G 2W1}

\author{Chris T.\ Bauch}
\affiliation{Department of Applied Mathematics, University of Waterloo, 200 University Avenue West, Waterloo, Ontario, Canada N2L 3G1}

\date{\today}

\begin{abstract}
Evolutionary games on graphs have been extensively studied. A variety of graph structures, graph dynamics, and behaviours of replicators have been explored. These models have primarily been studied in the framework of facilitation of cooperation, and much previous research has shed light on this field of study. However, there has been little attention devoted to truncation selection as most models employ proportional selection (such as in the replicator equation) or ``imitate the best.'' Here we systematically explore truncation selection on periodic square lattices, where replicators below a fitness threshold are culled and the reproduction probabilities are equal for all survivors. We employ two variations of this method: independent truncation, where the threshold is fixed; and dependent truncation, which is a generalization of ``imitate the best.'' Further, we explore the effects of diffusion in our networks in the following orders of operation: contest-diffusion-offspring (CDO), and diffusion-contest-offspring (DCO). CDO and DCO frequently facilitate and diminish cooperation, respectively. For independent truncation, we find three regimes determined by the fitness threshold: cooperation decreases as we raise the threshold; polymorphisms and extinction can occur; and the entire population goes extinct. Further, we show how an intermediate sucker's payoff maximizes cooperation in the DCO independent truncation model. We find that dependent truncation affects games differently; lower levels reduce cooperation for the Hawk Dove game and increase it for the Stag Hunt, and higher levels produce the opposite effects. We compare these truncation methods to proportional selection, and show that they can facilitate cooperation. We conclude that truncation selection can impact the prevalence of cooperation in complex ways, and therefore merit further study.
\end{abstract}

\maketitle


\section{\label{sec:level1}Introduction}

\begin{figure}
\includegraphics[]{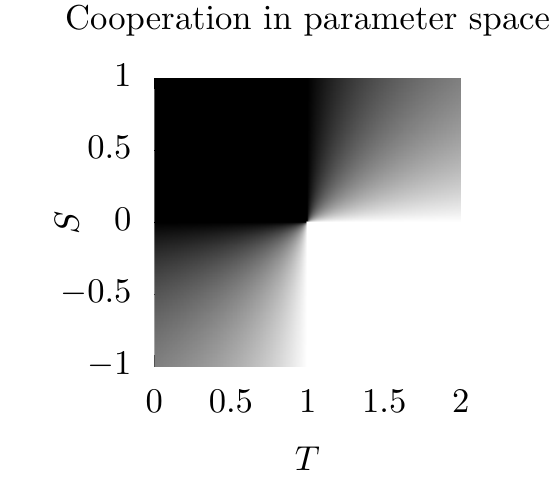}
\caption[Heatmap of cooperation in parameter space.]{Heatmap of cooperation in parameter space. White corresponds to defection and black to cooperation. In the harmony (top left quadrant), Hawk Dove (top right quadrant), and Prisoner's Dilemma (bottom right quadrant) areas of parameter space, the colour represents the equilibrium frequency of cooperation (white is All Defect, black is All Cooperate, and grey an intermediate amount of cooperation). In the Stag Hunt area (bottom left quadrant), we have bistability. Thus, the heatmap represents the magnitude of the basin of attraction of cooperation (i.e.\ $1 - \text{proportion of cooperators}$).}
\label{default}
\end{figure}

The evolution of cooperation is frequently modelled by the Prisoner's Dilemma. However, this model faces a tragedy of the commons in which defection is favoured over cooperation. The Prisoner's Dilemma is a symmetric game with two strategies: cooperate ($c$) and defect ($d$) with payoff matrix
\begin{equation}
  \mathbf{\Pi} = 
    \bordermatrix{ & c & d \cr
      c & R & S \cr
      d & T & P } \label{payoffMatrix}
\end{equation}

\noindent where $T > R > P > S$. Though the socially optimal strategy profile is $(c,c)$, due to the temptation to cheat, $T$, the suboptimal strategy profile $(d,d)$ is an evolutionarily stable state (ESS) \cite{hofbauer98}. This game and others are frequently studied in the parameter space determined by: $R = 1$, $-1 \leq S \leq 1$, $P = 0$, and $0 \leq T \leq 2$ \cite{santos06}. Thus, for $-1 \leq S < 0$ and $1 < T \leq 2$, we have the Prisoner's Dilemma. $0 < S \leq 1$ and $1 < T \leq 2$, we have the Hawk Dove game and a mixed ESS. For $-1 \leq S < 0$ and $0 \leq T < 1$, we have the Stag Hunt game and bistability. And, for $0 < S \leq 1$ and $0 \leq T < 1$, we have the harmony game where the ESS is socially optimal. Figure~\ref{default} depicts the frequency of cooperators at the interior equilibrium (if there is one) or at the exterior ESS. For the area of parameter space that represents the Hawk Dove game, this is the ESS. For the area of the Stag Hunt, this represents the size of the basin of attraction of cooperation.

A variety of evolutionary dynamics have been used to explore these games \cite{lehmann06,perc10}. Here we will focus on an extensively studied framework, evolutionary games on graphs \cite{nowak10,roca09,szabo07}, where agents are represented by vertices and interact with other vertices with which they share edges \cite{hauert01, nowak92}. Each vertex has a specific strategy that it follows. In many studies, including this one, we assume that the players always play the same pure strategy regardless of the strategies of their neighbours. From these interactions, vertices earn payoffs --- the average or sum of which is their fitness --- that determine survival by some selection method. Lattice models of this kind can increase cooperation \cite{killingback96} or reduce cooperation in the Snowdrift game \cite{hauert04}, depending on the selection method and game employed. A variety of different invasion scenarios have been studied in this framework \cite{fu10}, and it has been shown that cooperators can successfully invade \cite{langer08}.

Once payoffs have been assigned and fitnesses calculated, selection occurs. Each vertex compares its fitness to those of its neighbours to determine what strategy will occupy the vertex next turn. A common selection mechanism used in spatial games is proportional selection, where a vertex will randomly choose one of its neighbours, and adopt the neighbour's strategy with probability proportional to the difference in their fitnesses \cite{buesser12}, which is equivalent to the replicator equation for an infinite population \cite{helbing92}. Another common selection mechanism is ``imitate the best,'' in which the focal vertex will compare its fitness (the sum of all interactions with its neighbours) to the fitnesses of neighbouring vertices \cite{hauert01}. Its strategy will then become the strategy of the vertex with the greatest fitness. If there is a tie, it will be determined randomly from the maximal fitness neighbours.

Truncation selection occurs when a proportion of the population is culled and the survivors reproduce to fill the gap in the population. However, the reproduction rate is equal amongst all survivors (it is not scaled by fitness). Two types of truncation selection are dependent and independent \cite{morsky16}. In dependent truncation, the top $\tau$ of the population survives and reproduces, while the remainder are culled. For $\tau = 1$ on a graph, we have the ``imitate the best'' rule. In independent truncation, replicators with fitnesses greater than some fitness threshold $\phi$ survive and reproduce, and those below it are culled. Note that reproduction is not dependent upon the degree to which a replicator is above the threshold for survival. This asymmetry in selection results in significant differences from the replicator equation, potentially displaying chaos and significant levels of cooperation where none are represented in the replicator equation employing identical games \cite{ficici05, ficici07, fogel98, fogel11,morsky16}. A model that employs a degree of independent truncation is studied in \cite{zhang11}. Vertices are removed from the graph if their fitnesses are below a threshold. Vertices created as replacements have preferential connections to vertices with high fitnesses. After this process, there is proportional selection.

A variety of graphs have been explored, ranging from lattices with periodic or aperiodic boundaries, to small world graphs, and to random regular graphs \cite{buesser12}. Scale-free networks with different levels of degree-degree correlations and enhanced clustering have been shown to facilitate cooperation \cite{pusch08}. Cooperators perform better on random regular graphs than they do on regular small world graphs, which perform better than square lattices \cite{hauert05}.

Dynamic graphs are graphs where the edges change over time. By altering the degree of dynamism of the graph, a variety of mechanisms (such as the Red Queen) can lead to high levels of cooperation \cite{szolnoki09}. This process can be random or determined by vertex behaviour \cite{wardil14}. In the behavioural model, vertices may choose to break edges by examining the payoffs earned from neighbours with whom they share them \cite{cavaliere12,pacheco08}, breaking edges with non-cooperating neighbours \cite{rezaei12}, or form edges with those vertices that have high payoffs \cite{wu10,wu11}. Other means to study this behaviour include models where the agents move on a plane \cite{antonioni14, gomez07}. They interact with those within some Euclidean distance, which in some models is heterogeneous \cite{zhang11}. After a certain time they reproduce. Cooperation can be supported in such models, but only when the agents' velocities are low \cite{meloni09}. Scale-free graphs are the most resilient to this effect \cite{kun09}.

Another method of graph dynamism is diffusion, where vertices swap places in the graphs, or, equivalently, vertices swap strategies with neighbouring vertices. The order of the operations: contest, C; diffusion (graph dynamism), D; and offspring, O, heavily affects the results \cite{sicardi09,vainstein07}. CDO ordering of operations often inhibits the effects of graph structure \cite{sicardi09}.

Here we systematically explore independent and dependent truncation selection on periodic square lattices of a stochastic $2$ player game. Additionally, we study diffusion with both DCO and CDO operations. We compare our results to models that use Fermi selection --- which produces similar results to proportional selection \cite{buesser12} --- since it better accommodates stochastic payoffs.

\section{\label{sec:level2}Methods}

We construct a $50\times50$ periodic square lattice. We assign to vertices the cooperator strategy with probability $0.5$, and the remaining vertices are inhabited by defectors. We averaged $50$ realizations with $100$ turns each for each parameter value we explored, and employed synchronous contests, diffusion, and reproduction. The order of operations each turn is: contest, diffusion, and offspring for CDO; and diffusion, contest, and offspring for DCO, which we detail in the following paragraphs in the order: contest, diffusion, and offspring.

During the contest phase, players interact with each vertex with which they share an edge, and earn the payoff $\pi_{ij} + \xi$ from these interactions. $\xi \sim \mathcal{N}(0,0.4^2)$ 
is a Gaussian white noise variable and $\pi_{ij}$ is drawn from the payoff matrix,
\begin{equation}
 \mathbf{\Pi} = 
\begin{pmatrix}
1 & S \\
T & 0
\end{pmatrix}.
\end{equation}
\noindent From the payoffs received from each interaction we calculate the fitness of vertex $v_i$,
\begin{equation}
f_i = \frac{1}{|V_i|}\sum_{v_j \in V_i} \pi_{ij} + \xi,
\end{equation}
\noindent where $V_i$ is the set of inhabited neighbours of $v_i$ and $\pi_{ij}$ is the payoff $v_i$ receives playing neighbour $v_j$. We vary $T$ and $S$ with increments of $0.1$.

Diffusion occurs by randomly selecting vertices to swap strategies with their neighbours. Each turn we simulate $2500$ of these diffusion processes, which gives us an average diffusion rate of $1$. We explored higher diffusion rates ($d= 2, 5$). However, we found that these diffusion rates did not sufficiently affect our results.

During the offspring phase, the vertices' strategies are updated. We employ three different selection/updating rules: Fermi, dependent truncation, and independent truncation. For each vertex, we examine it and its neighbours' fitnesses and employ our selection method to determine what strategy will occupy the vertex next turn. For Fermi selection, we uniformly randomly select a vertex, $v_j$, from the set, $V_i \cup \{v_i\}$. Then, the probability that $v_i$ with strategy $s_i$ will adopt the neighbour's strategy, $s_j$, is
\begin{equation}
P(s_i \rightarrow s_j) = \frac{1}{1 + \exp(-(f_j-f_i)/\kappa)}.
\end{equation}
\noindent For $0.1 \leq \kappa \leq 1$. We explore the case where $\kappa = 1$.

For independent truncation we determine selection by the following algorithm: for each $v_i$ we determine the set
\begin{equation}
V_i^\prime = \{ v_j \in V_i \cup \{v_i\} : f_j \geq \phi \},
\end{equation}
\noindent where $\phi$ is the truncation threshold. We then set $v_i$'s strategy to the strategy of a randomly selected vertex of $V_i^\prime$. Note that independent truncation can result in $V_i^\prime = \emptyset$. In this case, $v_i$ becomes uninhabited. Such empty vertices hold no strategy and do not compete with their neighbours. However, they are still a part of the graph with respect to diffusion, and offspring may be born at them. If $|V_i| = \emptyset$, we set $f_i=0$ (as there are no opponents to play).

For dependent truncation, we compare the fitness of $v_i$ and its neighbouring vertices and determine the set of vertices that are in the top $\tau$ with respect to fitness. We then set the strategy of $v_i$ to the strategy of a randomly selected vertex from this set.

\section{\label{sec:level3}Results}

\subsection{Fermi selection}

\begin{figure}
\includegraphics[width=\linewidth]{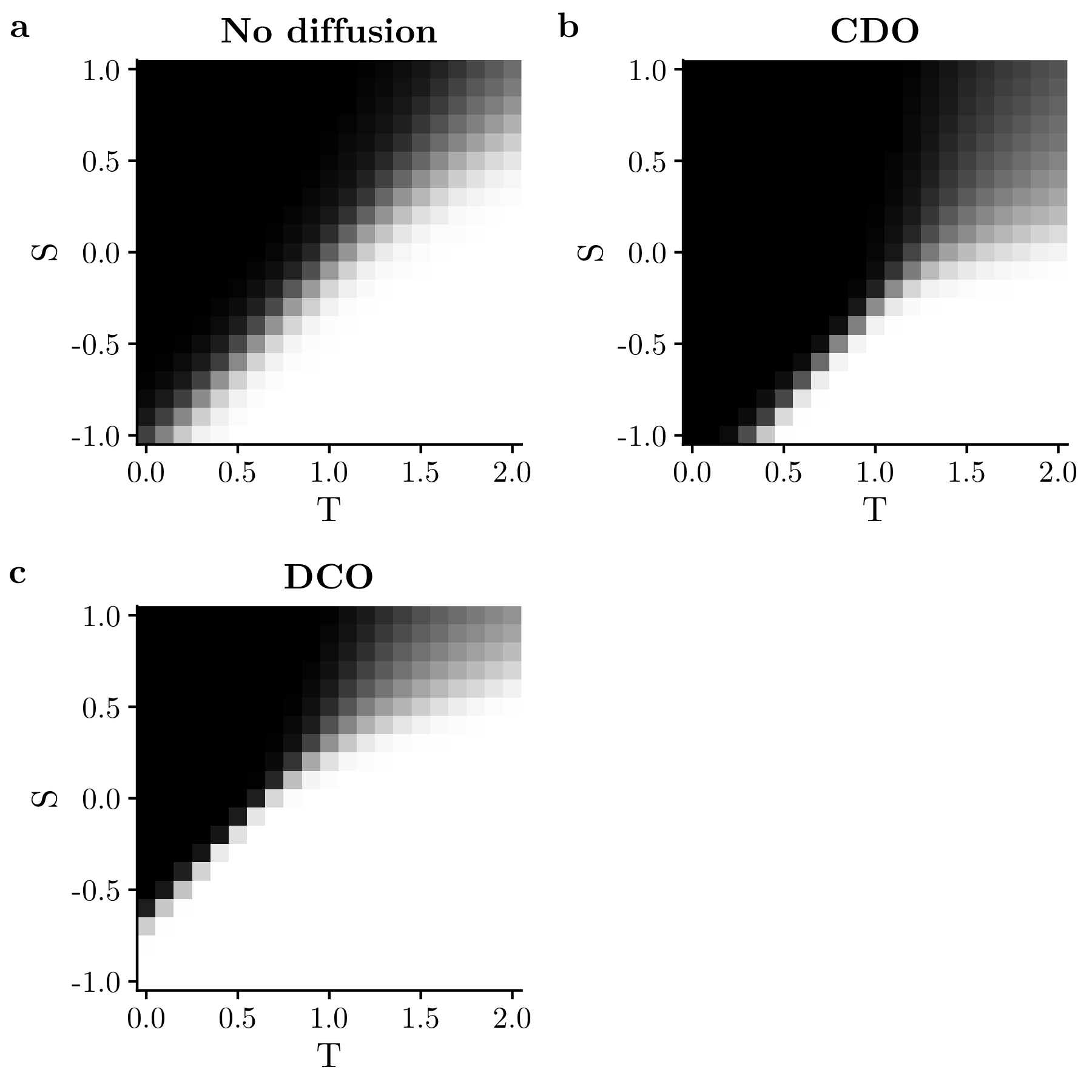}
\caption{Heatmap of average cooperation for Fermi selection with no diffusion (\textbf{a}), CDO (\textbf{b}), and DCO (\textbf{c}). White corresponds to defection and black to cooperation.}
\label{fermi}
\end{figure}

To enable comparison with the truncation selection results, in figure \ref{fermi} we plot the heatmap for Fermi selection with no diffusion, and for CDO and DCO. The no diffusion observation matches similar results in the literature, where clustering in structured models facilitates cooperation in the Stag Hunt game, and inhibits it in the Hawk Dove game \cite{roca09} relative to the mean field case (Figure \ref{default}). We observe less cooperation in Figure~\ref{fermi} than in Figure~\ref{default}, since the clustering coefficient, $\bar{C} = 0$, is low for our graph \cite{albert02}. CDO diffusion increases cooperation as observed in the Hawk Dove, Stag Hunt, and harmony domains of parameter space. Cooperators within a cluster of cooperators will have a higher fitness than defectors in a defector cluster. Thus, cooperators that diffuse into a defector cluster will propagate, whilst defectors that diffuse into a cooperator cluster will likely die. This effect has been similarly observed in \cite{sicardi09, vainstein07}. However, this phenomenon diminishes cooperation in the DCO case. Defectors can diffuse into a cluster of cooperators and compete with them thereby disrupting cooperation.

\subsection{Independent truncation}

We observe a variety of behaviours for independent truncation that can be classified into three regimes that are dependent upon the fitness threshold, $\phi$: $\phi \ll 0$, $0 \leq \phi \leq 1 $, and $\phi \gg 1$. In the first, the population does not go extinct and cooperation decreases as we increase $\phi$. In the second, cooperation can be facilitated and extinction primarily occurs in the Prisoner's Dilemma quadrant. And, in the third, extinction occurs everywhere in parameters space. Phase transitions with respect to $\phi$ are also observed in a mean field model of independent truncation \cite{morsky16}.

\begin{figure}
\includegraphics[width=\linewidth]{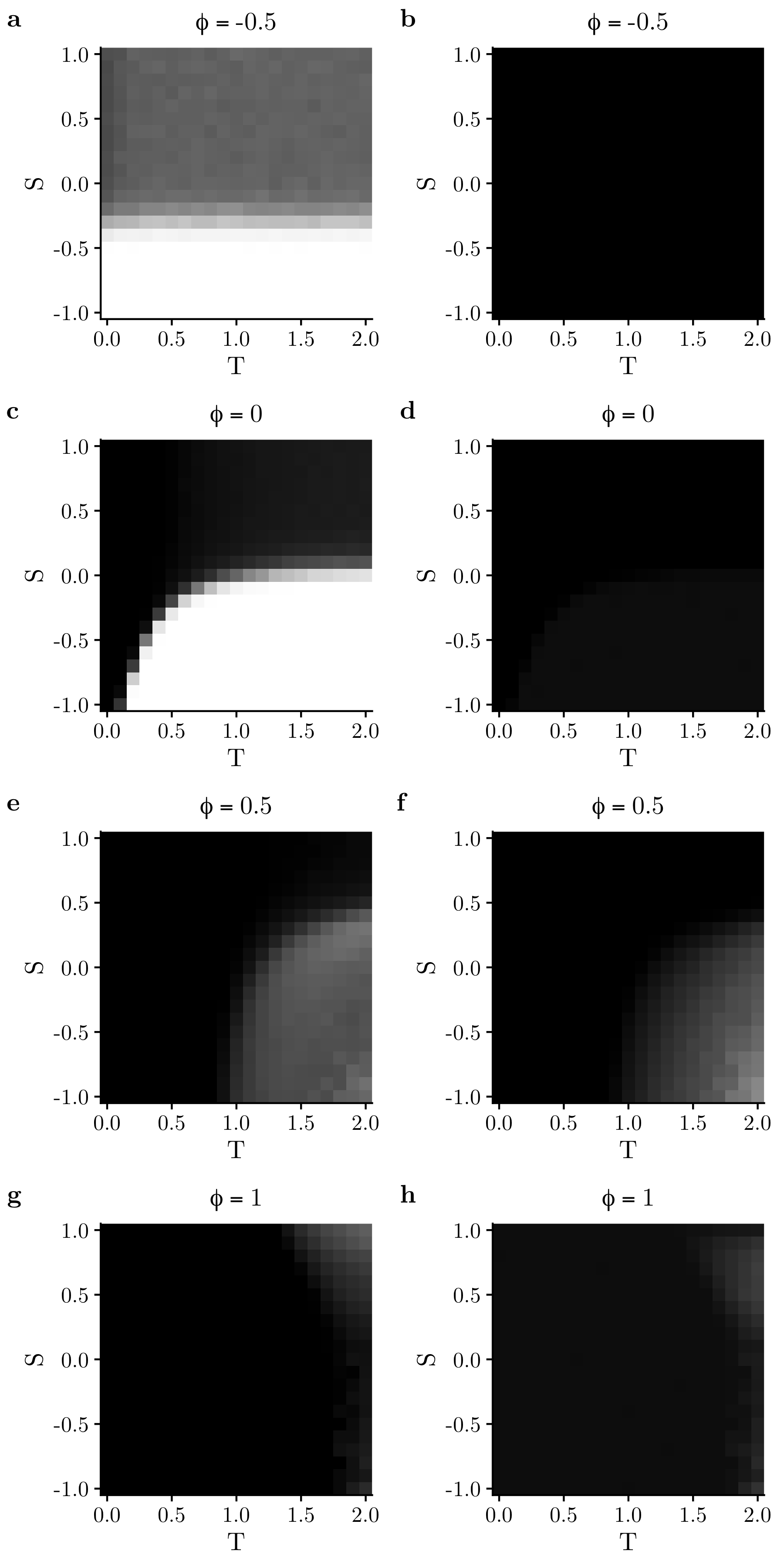}
\caption{Heatmap of average cooperation for independent truncation with no diffusion.  The left panels show the proportion of cooperators relative to the number of live vertices. White corresponds to defection and black to cooperation. The right panels show the proportion of live vertices. Black corresponds to live and white to empty vertices.}
\label{indepTrunc}
\end{figure}

Figure~\ref{indepTrunc} portrays a variety of interesting heatmaps of the independent truncation model with no diffusion. Panels \textbf{a} and \textbf{b} are representative of the results for $\phi \ll 0$. In this regime, defectors will frequently be above the threshold to survive throughout parameter space and regardless with whom they are playing. Cooperators, however, may be selected against if they compete with a sufficient number of defectors and $S$ is sufficiently low. For $S \gg \phi$, the population is approximately $50\%$ cooperator, since selection is effectively indifferent between the two strategies. However, as $\phi \to 0$, we observe a phase transition into the next regime.

Panels \textbf{c}-\textbf{h} of Figure \ref{indepTrunc} depict a qualitatively different phenomenon than $\phi \ll 0$. Here, selection will occur on both strategies. In panels \textbf{c} and \textbf{d}, we observe no extinction events and a greater degree of cooperation than when $\phi \ll 0$ and Fermi selection. 
Interestingly, when $\phi = 0.5$, we observe more cooperation than we do when $\phi = 0$. Note that the punishment payoff to defectors is zero ($P = 0$). At $\phi = 0$, this means that approximately half of defectors will frequently be selected against when playing many of their own strategy, and half will not. However, at $\phi = 0.5$, the deleterious effect to a defector playing against defector neighbours is much higher. Thus, defector clusters are less stable than at a lower $\phi$. This phenomenon gives cooperators an advantage and allows them to better flourish. Seeing that $R = 1$, cooperative clusters are stable. Further, they may sustain defectors ($T > 1$). These phenomena sustain a polymorphic population in the Prisoner's Dilemma quadrant for $\phi = 0.5$ (\textbf{e} and \textbf{f}). However, for $\phi = 1$ (\textbf{g} and \textbf{h}), polymorphisms are more prevalent in the Hawk Dove quadrant. This phenomenon is more pronounced in the DCO model. 
For $\phi \gg 1$, cooperation cannot be sustained and extinction occurs throughout parameter space (results not shown here).

\begin{figure}
\includegraphics[width=\linewidth]{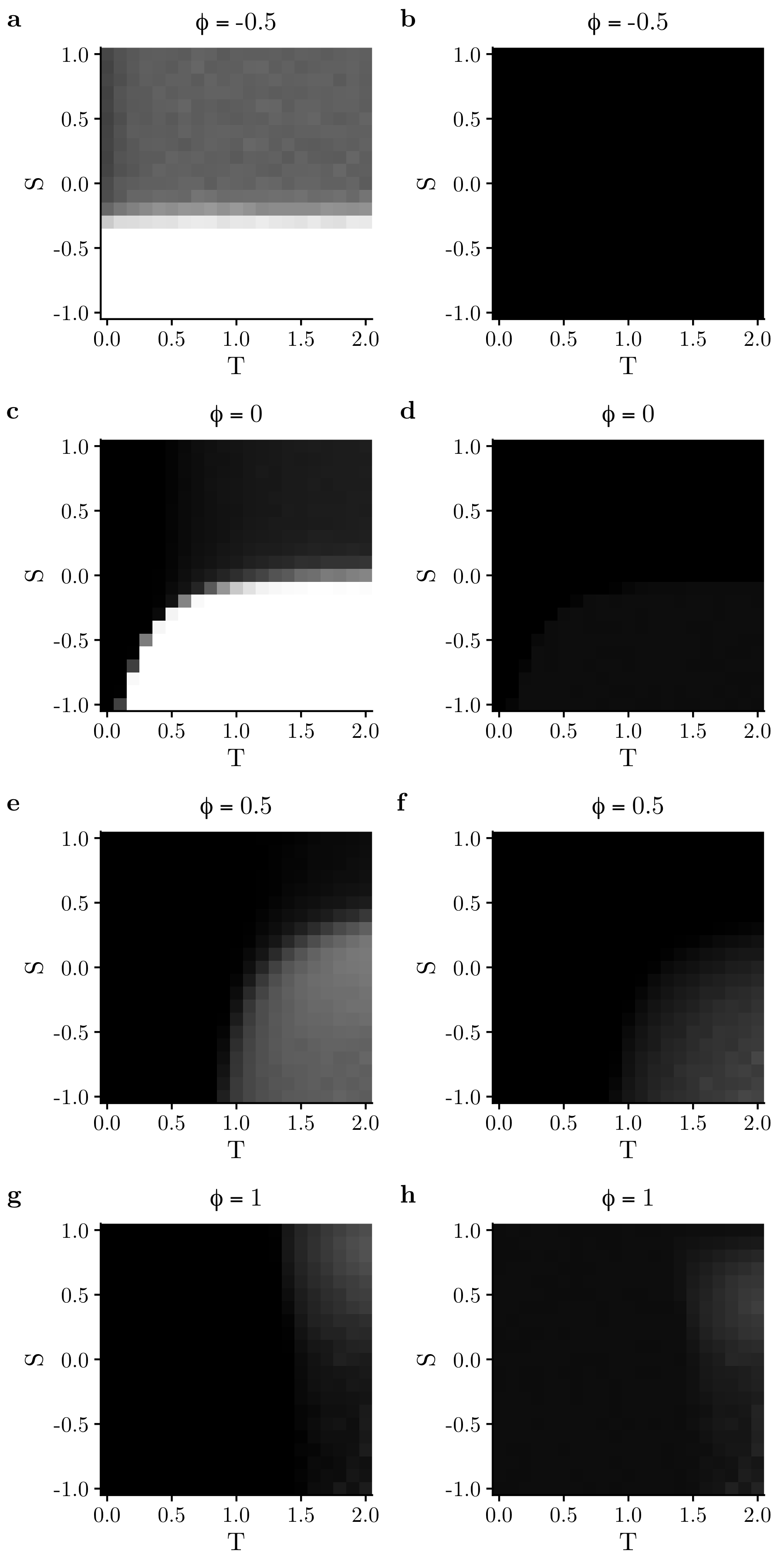}
\caption{Heatmap of average cooperation for independent truncation (CDO).  The left panels show the proportion of cooperators relative to the number of live vertices. White corresponds to defection and black to cooperation. The right panels show the proportion of live vertices. Black corresponds to live and white to empty vertices.}
\label{indepTruncCDO}
\end{figure}

Figure \ref{indepTruncCDO} depicts the results for the CDO model. These results are remarkably similar to the model without diffusion (Figure \ref{indepTrunc}). The higher fitness a cooperator may receive from being in a cooperator cluster does not benefit them if they diffuse into a defector cluster, since we are selecting for adequacy. Thus, unlike our other models, CDO does not appreciably increase cooperation with respect to no diffusion.



\begin{figure}
\includegraphics[width=\linewidth]{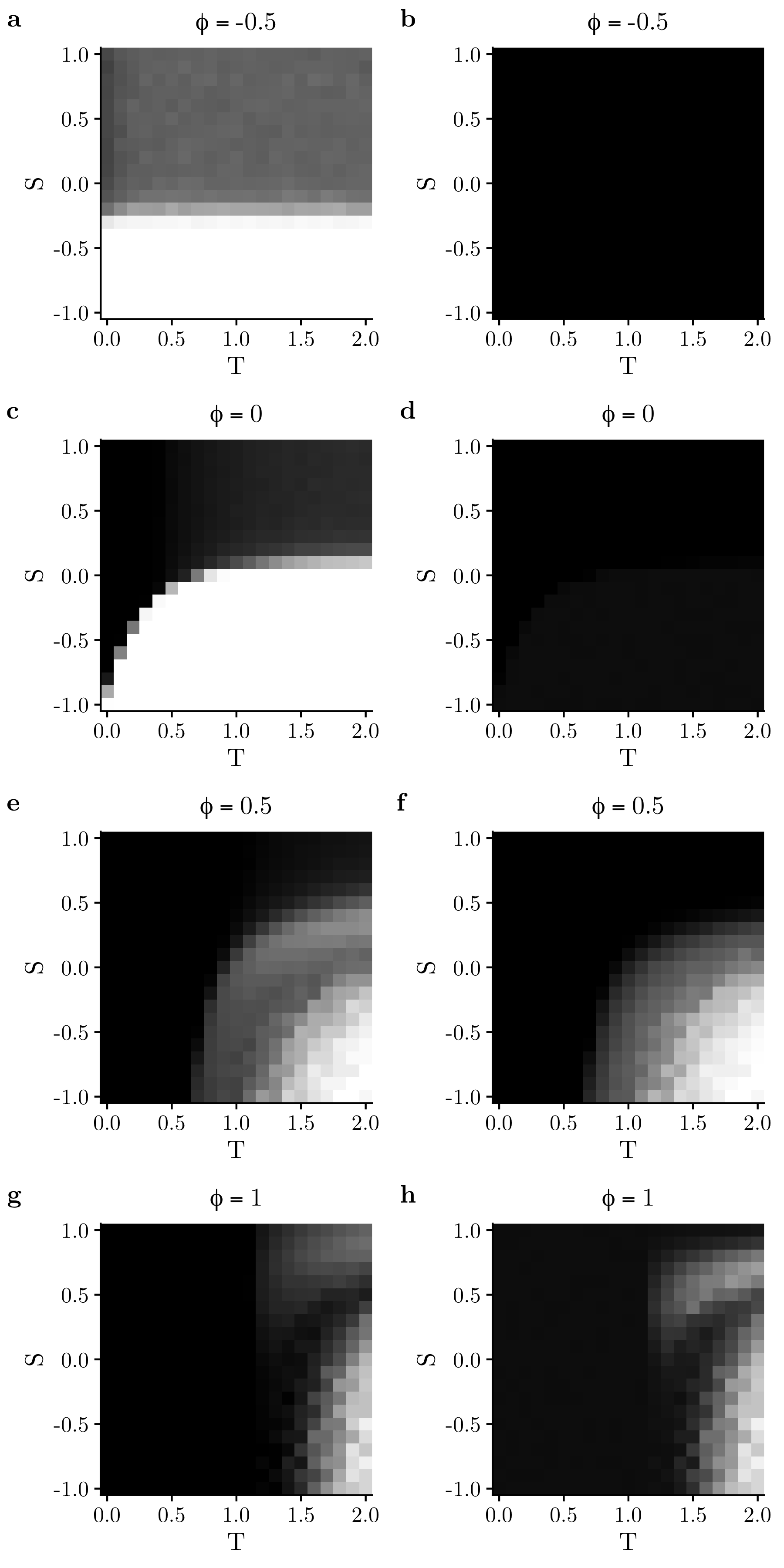}
\caption{Heatmap of average cooperation for independent truncation (DCO). The left panels show the proportion of cooperators relative to the number of live vertices. White corresponds to defection and black to cooperation. The right panels show the proportion of live vertices. Black corresponds to live and white to empty vertices.}
\label{indepTruncDCO}
\end{figure}

Unlike the CDO model, the DCO model of independent truncation differs from the no diffusion case for appropriate $\phi$ as depicted in Figure \ref{indepTruncDCO}. We do not observe differences when $\phi = -0.5$ (\textbf{a} and \textbf{b}). For $\phi = 0$, we observe less cooperation for DCO, which aligns with our results for Fermi selection and dependent selection. Diffusing defectors can push the fitnesses of cooperators in a cooperator cluster below the threshold. This effect, for $\phi = 0.5$ (\textbf{e} and \textbf{f}), results in extinction and less cooperation. Defectors are able to disrupt cooperation by diffusing into cooperative clusters. And, in some areas of parameter space, drive cooperators to extinction. However, without cooperators to sustain defectors, they also go extinct. This effect is greatest in the Prisoner's Dilemma region, but also affects parameter space bordering it.


\begin{figure}
\includegraphics[width=\linewidth]{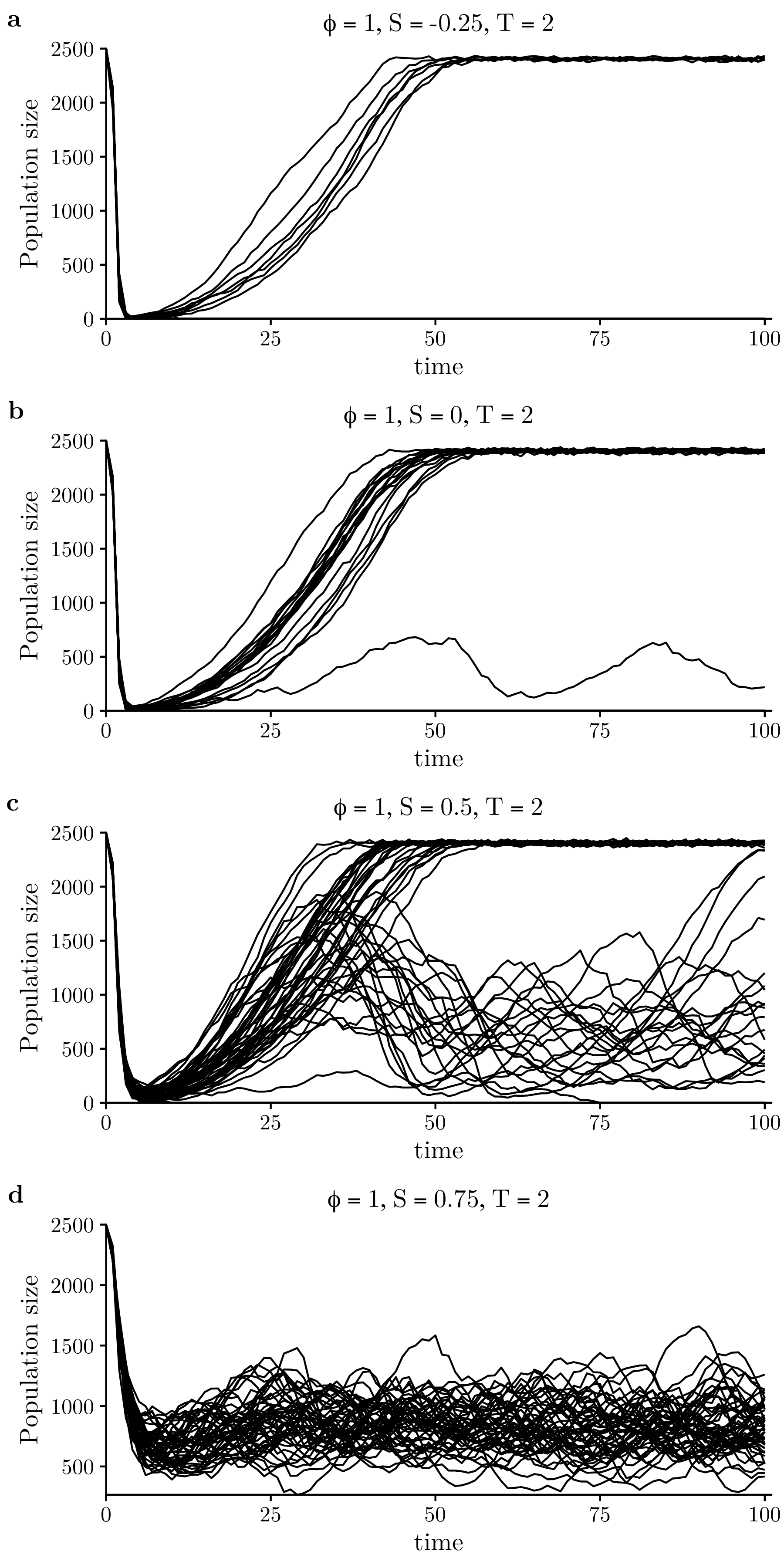}
\caption{Time series for the DCO independent truncation model. Each panel plots the total population for $50$ time series for a random initial condition. Cooperation is optimized for an intermediate value of $S$ (panel \textbf{c}). If $S$ is too low, extinction of both strategies is likely (panels \textbf{a} and \textbf{b}). However, if $S$ is too high (panel \textbf{d}), defectors can be sustained, thereby restraining cooperation and the total population size.}
\label{time_series}
\end{figure}

For $\phi = 1$, we observe an interesting phenomenon where there is an intermediate optimal $S$ with respect to cooperation when $T$ is high (Figure \ref{indepTruncDCO} \textbf{g} and \textbf{h}). Figure \ref{time_series} depicts the time series so as to explain these results. When $S$ is too low (Figure \ref{time_series} \textbf{a}), defectors frequently grow too quickly and cause the population as a whole to crash, and polymorphisms are unstable. As we increase $S$, the basin of attraction to the all cooperator state increases. The population initially crashes. And, though defectors become extinct, cooperators can often survive and grow to near the carrying capacity (Figure \ref{time_series} \textbf{b}). For $T = 2$, cooperation is optimally facilitated at $S \approx 0.5$ (Figure \ref{time_series} \textbf{c}). Extinction is rare, and complete cooperation is often reached. Paradoxically, at a higher $S$ (Figure \ref{time_series} \textbf{d}), cooperation is diminished. Here, the initial population crash is less severe and a polymorphism is more stable.

\subsection{Dependent truncation}

\begin{figure}
\includegraphics[width=\linewidth]{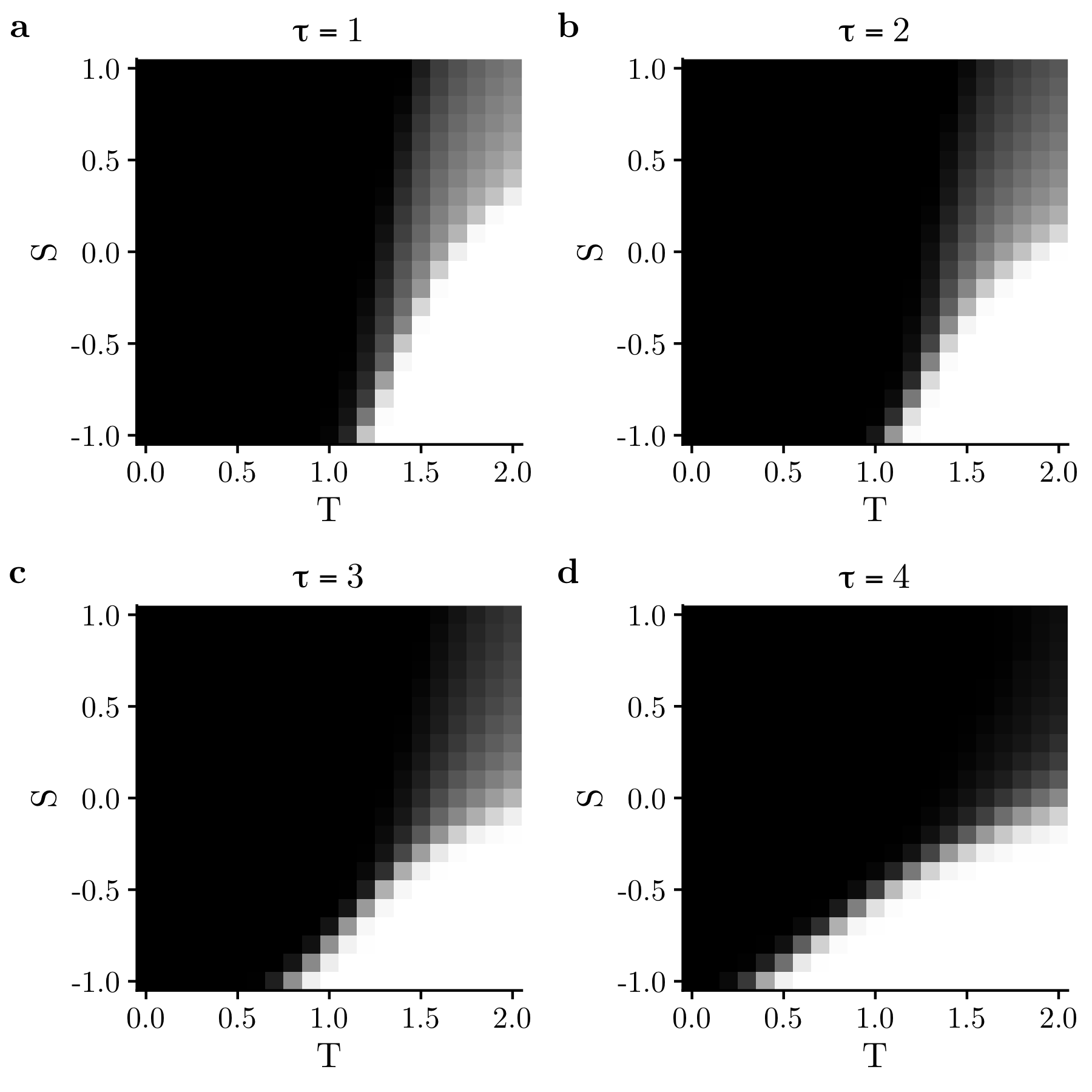}
\caption{Heatmap of average cooperation for dependent truncation with no diffusion. White corresponds to defection and black to cooperation.}
\label{depTrunc}
\end{figure}

\begin{figure}
\includegraphics[width=\linewidth]{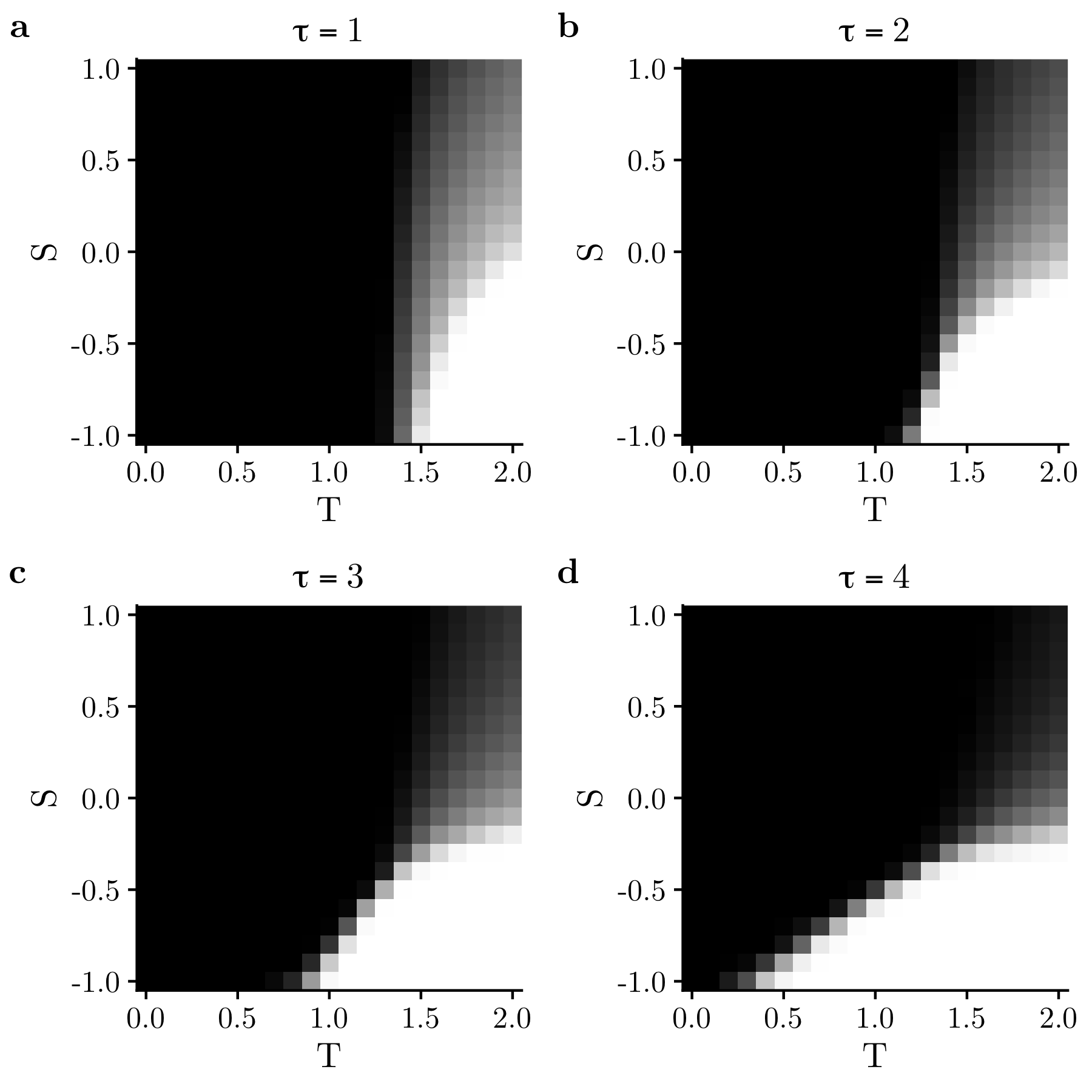}
\caption{Heatmap of average cooperation for dependent truncation (CDO). White corresponds to defection and black to cooperation.}
\label{depTruncCDO}
\end{figure}

\begin{figure}
\includegraphics[width=\linewidth]{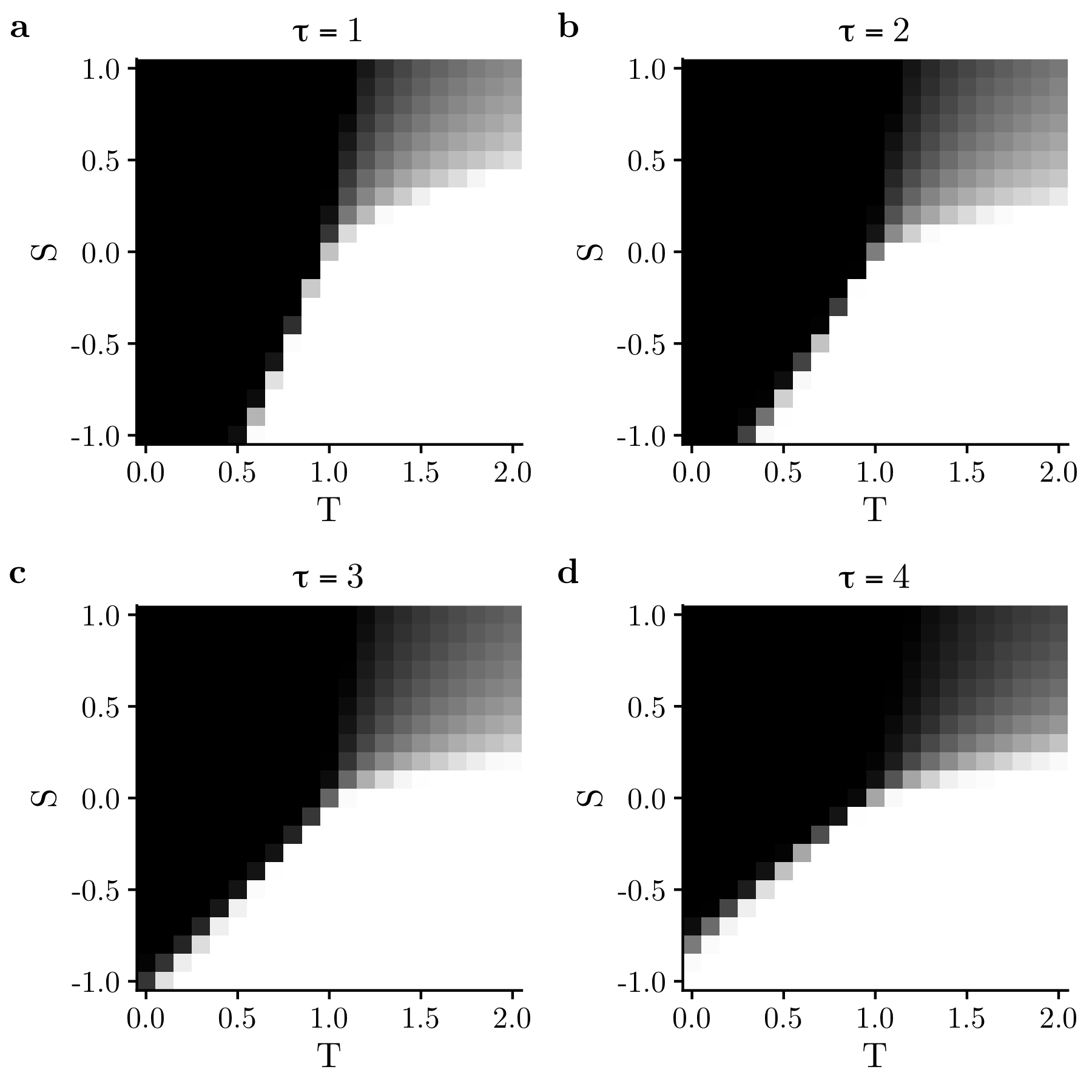}
\caption{Heatmap of average cooperation for dependent truncation (DCO). White corresponds to defection and black to cooperation.}
\label{depTruncDCO}
\end{figure}

In general, for a given diffusion model, we observe more cooperation in dependent truncation than we do in Fermi selection regardless of the truncation level, $\tau$. This pattern holds for no diffusion, CDO, and DCO, as shown in Figures \ref{depTrunc}, \ref{depTruncCDO}, and \ref{depTruncDCO}, respectively. However, the value of $\tau$ affects each game differently, which can be seen by examining the quadrants of the figures.

Figure~\ref{depTrunc} depicts the heatmap results for dependent truncation with no diffusion for $\tau = 1, 2, 3, 4$. Looking to the top left quadrant of the figures, we see that there is no effect of varying $\tau$ upon the harmony game; all players cooperate. However, the levels of cooperation decrease in the Stag Hunt game as $\tau$ is increased. The defection strategy in the Stag Hunt game can be interpreted as a risk averse strategy in that it always guarantees (not including the stochastic component) a payoff between the maximum and minimum that may be received by a cooperator. If $\tau$ is low, then survival this risk averse strategy is unfit, since the model is selecting for high fitness. However, for a higher $\tau$, the model selects against the lowest fitness. In such a case, a risk averse strategy that avoids the sucker's payoff, $S$, can be beneficial. Nonetheless, even when $\tau = 4$, we still have higher cooperation in the Stag Hunt game for dependent truncation than we do for Fermi selection (Figure \ref{fermi}).

While $\tau$ is negatively correlated with cooperation in the Stag Hunt, it is positively correlated with it in the Hawk Dove game. As in the Stage Hunt, an understanding of this phenomenon can be found in considering $\tau$ as selecting for high fitness or against low fitness. Hawks (defectors) have the higher variance in fitness unlike the stag hunters (cooperators). Thus, when $\tau$ is low, selection is for high fitness individuals, which will more likely be defectors in the Hawk Dove game where they were cooperators in the Stag Hunt. Likewise, a high $\tau$ will select for low variance players, the doves (cooperators).

There is little effect upon the low levels of cooperation for the Prisoner's Dilemma. However, $\tau$ affects regions of the Prisoner's Dilemma quadrant differently. Close to the boundary with the Stag Hunt quadrant, increasing $\tau$ inhibits cooperation (as it does for the Stag Hunt game). Similarly, near the boundary with the Hawk Dove game, varying $\tau$ exhibits the same effect as it does in the Hawk Dove game; it is correlated with cooperation.

Figures \ref{depTruncCDO} and \ref{depTruncDCO} mirror the qualitative observations we have made thus far. $\tau$ has affects each game the same as it does without diffusion. Further, we observe a similar phenomenon as in Fermi selection regarding the algorithm's order of operation. CDO facilitates cooperation and DCO inhibits it. This result is insensitive to $\tau$. In particular, the Prisoner's Dilemma quadrant is devoid of cooperation in DCO, but is present in CDO.

\section{Discussion}

Here we have systematically explored diffusion and different selection mechanisms on a square lattice of cooperators and defectors. We have expanded the understanding of selection mechanisms --- such as Fermi selection and ``imitate the best'' --- to incorporate stochastic payoffs, two truncation schemes, various levels of truncation, and two diffusion algorithms.

We have uncovered three regimes for independent truncation: cooperation decreases as we increase the threshold parameter, $\phi$; polymorphisms and extinction can occur; and extinction of the total population occurs. For the DCO model of independent truncation, we observe that for a high temptation, cooperation is best facilitated if the sucker's payoff is intermediate. We can understand this result by considering the cooperators as prey and defectors as predators. The sucker's payoff is therefore the death rate due to predation. At our initial conditions, the population experiences an initial crash, which drives the defectors to extinction. If defectors survive the initial crash, the population faces oscillations of high amplitude, which results in further population crashes. If the sucker's payoff is too low, cooperators often cannot survive a crash and thus the whole population of replicators goes extinct. As the sucker's payoff is increased, cooperators are better able to survive a crash, and thus we see the system frequently evolving to being wholly composed of cooperators. However, if the sucker's payoff is too high, then the population crash is mitigated, which prevents defectors from going extinct. We thereby reach a stable polymorphism and the amplitude of the oscillation in the total population are relatively small.

Dependent truncation is an extension of ``imitate the best.'' However, we vary how many of the best players from which we choose for reproduction (by choosing $\tau$). For low $\tau$, we select from the very best of the population. For high $\tau$, the majority of the players may be chosen to reproduce. And, as such, we are selecting against the least fit players. $\tau$ has different effects on the density of cooperators for different games. While the harmony and prisoner's dilemma were not much affected, the Hawk Dove and Stag Hunt were. As we raise $\tau$, we decrease cooperation in the Stag Hunt game, but increase it in the Hawk Dove game. This phenomenon occurs with and without diffusion (and for both DCO and CDO).

Both of these truncation methods can have substantially more cooperation than observed for Fermi selection given the same diffusion rule. However, independent truncation can have less cooperation than Fermi selection when the threshold, $\phi$, is low.

The impact of diffusion is most acute for one diffusion event per player on average, $d=1$. We ran simulations up to $d=5$ and observed only negligible effects upon our results. The DCO algorithm permits the disruption of clusters of cooperators by allowing defectors on the edge of the cluster to diffuse into it, where they exploit and out-compete their cooperating neighbours (though this effect is muted in independent truncation). In the CDO algorithm, however, players play their neighbours and then diffuse. Thus, cooperators in clusters earn high fitnesses and then may diffuse into defector clusters, where fitnesses are low. Defectors that diffuse into cooperative clusters will have low fitnesses earned from their defector neighbours, and thus cannot become established within the cooperative clusters and thereby disrupt them.

While a variety of graphs and diffusion processes have been explored in the literature, truncation selection on evolutionary graphs has not. The generalization of ``imitate the best'' and the rich behaviours these models can models can produce suggest much interesting future research.

\bibliographystyle{elsarticle-harv}
\bibliography{bib_paper3}

\end{document}